\documentclass[twocolumn,preprintnumbers,amsmath,amssymb,prl]{revtex4}

\usepackage{graphicx}
\usepackage{dcolumn}
\usepackage{bm}

\begin{document}

\title{Coherent Population Trapping of Single Spins in Diamond Under
    Optical Excitation}

\author{Charles Santori$^1$}
 \email{charles.santori@hp.com}
\author{Philippe Tamarat$^2$}
\author{Philipp Neumann$^2$}
\author{J\"{o}rg Wrachtrup$^2$}
\author{David Fattal$^1$}
\author{Raymond G. Beausoleil$^1$}
\author{James Rabeau$^3$}
\author{Paolo Olivero$^3$}
\author{Andrew D. Greentree$^{3,4}$}
\author{Steven Prawer$^{3,4}$}
\author{Fedor Jelezko$^2$}
\author{Philip Hemmer$^5$}

\affiliation{$^1$Hewlett-Packard Laboratories,
             1501 Page Mill Rd., Palo Alto, CA 94304, USA}
\affiliation{$^2$ 3. Physikalisches Institut, Universit\"{a}t Stuttgart, 70550 Stuttgart, Germany}
\affiliation{$^3$ School of Physics, University of Melbourne, Victoria, Australia, 3010}
\affiliation{$^4$ Centre of Excellence for Quantum Computer Technology, School of Physics, University of Melbourne, Victoria, Australia, 3010}
\affiliation{$^5$ Electrical \& Computer Engineering Department, Texas A\&M University, College Station, Texas 77843, USA}

\begin{abstract}
Coherent population trapping is demonstrated in single nitrogen-vacancy centers in diamond under optical excitation. For sufficient excitation power, the fluorescence intensity drops almost to the background level when the laser modulation frequency matches the 2.88 GHz splitting of the ground states. The results are well described theoretically by a four-level model, allowing the relative transition strengths to be determined for individual centers. The results show that all-optical control of single spins is possible in diamond.
\end{abstract}

\date{\today}
\maketitle

Using optical laser fields to manipulate single spins in solids is a promising path toward solid-state quantum information processing.  An important advantage of this technique over direct microwave excitation of spin transitions is micron-scale spatial resolution, which enables selective addressing of individual qubits~\cite{shahriar02}.  Optical spin control is also important for interfacing flying and stationary qubits as needed for quantum networks~\cite{cirac97} and repeaters~\cite{childress05}.

Closely related to this is the effect known as coherent population trapping (CPT)~\cite{arimondo76}, observed first in gasses~\cite{gray78} and later developed into electromagnetically induced transparency (EIT)~\cite{boller91}.  When multiple spin levels are driven by optical fields to a common excited state (a $\Lambda$ configuration), a non-absorption resonance can occur due to destructive quantum interference between two absorption pathways. A dark state forms which is a coherent superposition of two ground states with probability amplitudes tunable through the laser amplitudes.  CPT can be viewed as a steady-state version of optical spin control, while time-varying fields allow for dynamic control.  An important requirement is long-lived ground-state spin coherence even under strong optical excitation of the material.

Coherent population trapping and EIT have now been obtained in a variety of solids. For example, extremely long storage times and room-temperature CPT have been achieved in Pr:YSO~\cite{turukhin02,longdell05} and ruby~\cite{kolesov05}, respectively, but the oscillator strengths in these materials seem too small for experiments with single impurities.  In semiconductor systems the oscillator strength can exceed unity, and in single charged quantum dots optical pumping~\cite{atature06} and initialization of a particular coherent superposition of spin states~\cite{dutt05} have recently been reported, representing a promising step toward all-optical spin control. However, in quantum dots as well as shallow donors~\cite{fu05} the spin coherence is thought to be limited by hyperfine interaction with randomly oriented nuclear spins.  Another promising system is the nitrogen-vacancy (NV) defect in diamond, which has been identified as a promising qubit because of its long phase memory~\cite{kennedy03}. Here, we show that by isolating a single NV center in diamond we can obtain a nearly ideal CPT resonance.

Composed of a substitutional nitrogen next to a carbon vacancy, the NV center can have extremely long-lived spin coherence because the diamond lattice is composed primarily of $^{12}$C, which has zero nuclear spin.  The NV center has been studied extensively with recent emphasis on potential applications in quantum information processing. For the negatively-charged NV center, the ground states (Fig. 1) consist of a spin triplet with a $2.88\,\mathrm{GHz}$ splitting between the lower $m_{s}=0$ level and the upper $m_{s} =\pm 1$ levels~\cite{vanoort88}. These are connected to excited states by optical transitions of moderate strength, with a total oscillator strength of approximately $0.2$ for the total vibronic band, or $0.006$ for the zero-phonon line alone.  By exciting these transitions and detecting the resulting fluorescence, readout of single spins~\cite{jelezko02} and optically-detected electron spin resonance in a single NV center~\cite{jelezko04a} have been demonstrated. This has been extended to controlled coupling between the electronic spin of a single NV center and a nearby electronic or nuclear spin~\cite{gaebel06,hanson06,jelezko04b}.
\begin{figure}[tb]
\includegraphics[]{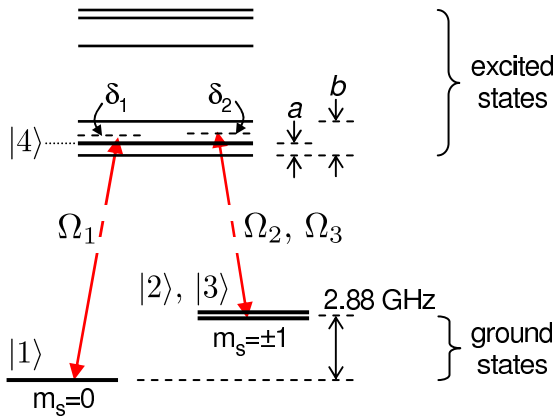}
\caption{\label{fig1} Schematic energy level diagram of a strained NV center showing resonant optical frequencies coupling ground states $|1\rangle$, $|2\rangle$ and $|3\rangle$ to excited state $|4\rangle$. $\Omega_i$ are the Rabi frequencies, $\delta_i$ are the optical detunings, and $a$ and $b$ are the splittings in the lowest three excited states.}
\end{figure}

For NV centers it is not clear in the current literature how best to realize a $\Lambda$ system due to uncertainty in the excited-state structure.  Spectral-hole-burning studies on large ensembles indicate excited levels which sometimes couple to multiple ground sublevels~\cite{reddy87,manson94}, while recent work on single centers suggests parallel, spin-conserving transitions at zero magnetic field~\cite{jelezko02}.  One approach for obtaining a $\Lambda$ system is to use a magnetic field to mix the ground states~\cite{hemmer01}. Alternatively, a $\Lambda$ system can be found even at zero magnetic field due to factors such as strain which reduce the symmetry and primarily modify the excited states~\cite{santori06}.  The level diagram in Fig. 1 is appropriate for the strained case.  The excited states are split by their orbital components into two spin triplets.  As discussed in Ref.~\cite{manson06}, under strain the non-spin-conserving transitions can be enhanced through the spin-orbit interaction.
\begin{figure}[tb]
\includegraphics[]{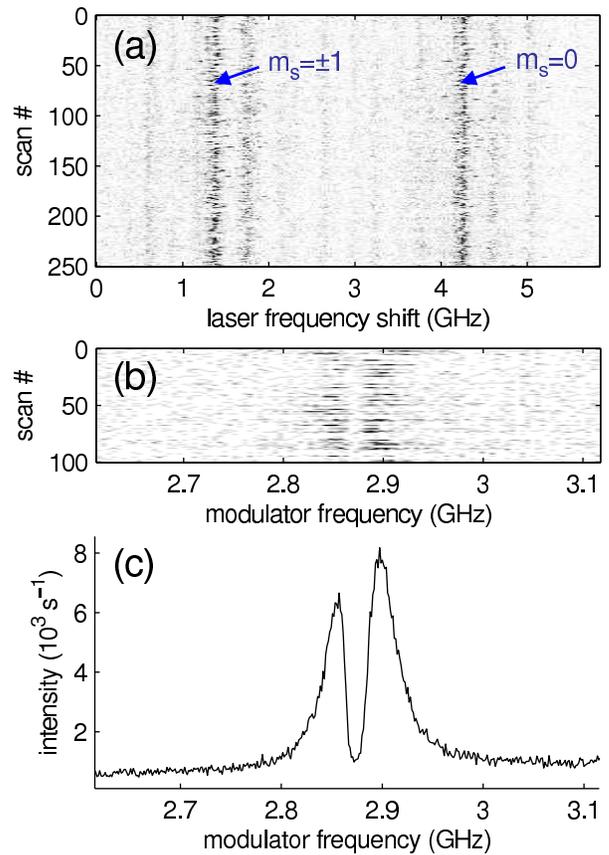}
\caption{\label{fig2} Experiments at $T = 8 \, \mathrm{K}$ with laser modulation sidebands approximately 50\% as intense as the fundmental component:
(a) Fluorescence intensity recorded as the laser frequency was scanned repeatedly while the modulator was fixed at $2.90 \,\mathrm{GHz}$.  The two bright lines are from a single NV in the center of the field, while most of the weak lines are from background NVs.
(b) Repeated modulator frequency scans with laser fixed on $m_s=0$ transition.
(c) Sum of many modulator scans, including only those with total fluorescence intensity above a threshold.}
\end{figure}

We performed measurements on natural type IIa diamond samples obtained from URAL with a $(111)$ orientated top surface.  The samples have low enough NV concentrations that single centers can be accessed individually by confocal microscopy~\cite{tamarat06}. The samples were cooled to $2-10\,\mathrm{K}$ in a liquid-helium cryostat. To increase the proportion of NV centers with allowed spin-flip transitions, we applied stress mechanically by mounting the sample such that it was squeezed by the copper sample mount when it contracted upon cooling. A piezoelectric transducer allowed for additional adjustments. By this method we obtained a splitting in the ensemble zero-phonon line of $\sim 30\,\mathrm{GHz}$ corresponding to a stress of $\sim 30\,\mathrm{MPa}$.  Under this condition we could find a spectral component for which most of the NV centers had allowed spin-flip transitions.  However, because of strain and disorder naturally present in many samples, externally applied stress is not always necessary to find NV centers with allowed spin-flip transitions. The magnetic-field measurements described below were performed without externally applied stress.

The main steps in the experiment are shown in Fig.~2.  An excitation laser, resonant with the $637\,\mathrm{nm}$ zero-phonon line of the negatively-charged NV center, was focused to a diffraction-limited spot within the sample. Detection of fluorescence into the phonon sidebands from approximately $670-750\,\mathrm{nm}$ provided a measure of excited-state population.  Since fluorescence from NV centers with allowed spin-flip transitions cannot be observed with a single excitation frequency due to optical pumping, we used an electro-optic modulator to produce sidebands at $\pm 2.90 \,\mathrm{GHz}$, slightly off resonance from the ground-state splitting, to excite all ground states simultaneously.  When the laser frequency is scanned, two bright fluorescence peaks belonging to the same center are observed (Fig.~2a).  For the left peak, the laser frequency excites the $m_{s}=\pm 1$ ground states while a modulation sideband excites the $m_{s}=0$ state. For the right peak, the laser excites the $m_{s}=0$ state while the other sideband excites the $m_{s} =\pm 1$ states. Between scans, a pulsed repump laser ($532\,\mathrm{nm}$) was applied to re-initialize the NV center.  This was required because the NV center was observed to bleach after $\sim 10^6$ fluorescence cycles of resonant excitation.  A crucial feature of these diamond samples is that single NV centers often exhibit spectral diffusion over a frequency range of $100\,\mathrm{MHz}$ or less even after the repump is applied as in Fig.~2a.

To observe coherent population trapping, the laser frequency was fixed on one of the peaks, and the modulation frequency was scanned across $2.88 \,\mathrm{GHz}$ with a $10\,\mathrm{Hz}$ repetition rate (Fig.~2b).  Summing over many scans reveals a broad peak centered at $2.88 \,\mathrm{GHz}$ with a narrow, central dip extending almost down to the background level, suggesting a large degree of spin coherence on two-photon resonance.  The blinking between scans is due to the stochastic nature of the re-pump process.   For the data in Fig.2c, a threshold procedure was used to include only those scans (about 50\%) for which the NV center was in its optically active state. This reduces the background level with little effect on the shapes of the curves.

To understand this behavior theoretically, we use a four-level model which includes the three ground states $|1\rangle$ , $|2\rangle$ and $|3\rangle$ representing $m_{s} =0$ and two orthogonal linear combinations of $m_{s} =\pm 1$, and an excited state $|4\rangle$. The effective Hamiltonian in a rotating frame under the rotating-wave approximation is,
\begin{equation}
\frac{H}{\hbar } =\left(
\begin{array}{cccc}
\delta _{1} & 0 & 0 & \Omega _{1}^{\ast } /2 \\
0 & \delta _{2} & 0 & \Omega _{2}^{\ast } /2 \\
0 & 0 & \delta _{2} +\delta _{23} & \Omega _{3}^{\ast } /2 \\
\Omega _{1} /2 & \Omega _{2} /2 & \Omega _{3} /2 & 0%
\end{array}
\right) ,
\end{equation}
where $\delta_{1}$ and $\delta_{2}$ are the laser frequency detunings from the 1-4 and 2-4 transitions, respectively, $\delta_{23}$ is the level 2-3 splitting, and $\Omega_{i} $ are the Rabi frequencies proportional to the square root of the laser intensities. When spontaneous emission is also taken into consideration, only eigenstates with zero probability amplitude in $|4\rangle$ are stable (dark states), and all other states decay through fluorescence. Dark states can occur under two conditions: $\Omega_{2} |1\rangle - \Omega_{1} |2\rangle$ is a dark state if $\delta_{1} =\delta_{2}$, and $\Omega_{3} |1\rangle -\Omega_{1} |3\rangle$ is a dark state if $\delta_{1} =\delta_{2} +\delta_{23}$. For the degenerate case $\delta_{23}=0$ we have a two-dimensional dark subspace.  To obtain a quantitative prediction for the fluorescence intensity, which is proportional to the excited-state population, we must include spontaneous emission from $|4\rangle$ as well as additional decoherence terms.  We then solve a master equation~\cite{cohentannoudi} for the steady-state density matrix $\rho$ according to,
\begin{equation}
d\rho /dt=-i[H/\hbar,\rho \rbrack +{\mathcal R}[\rho \rbrack =0 \, .
\end{equation}
For ${\mathcal R}[\rho \rbrack $ we included population decay rates
$\Gamma_{i}$ from $|4\rangle$ to ground states $|i\rangle$ and decay
rates $\gamma_{ij}$ of the off-diagonal density matrix elements
$\rho_{ij}$.

\begin{figure}[tb]
\includegraphics[]{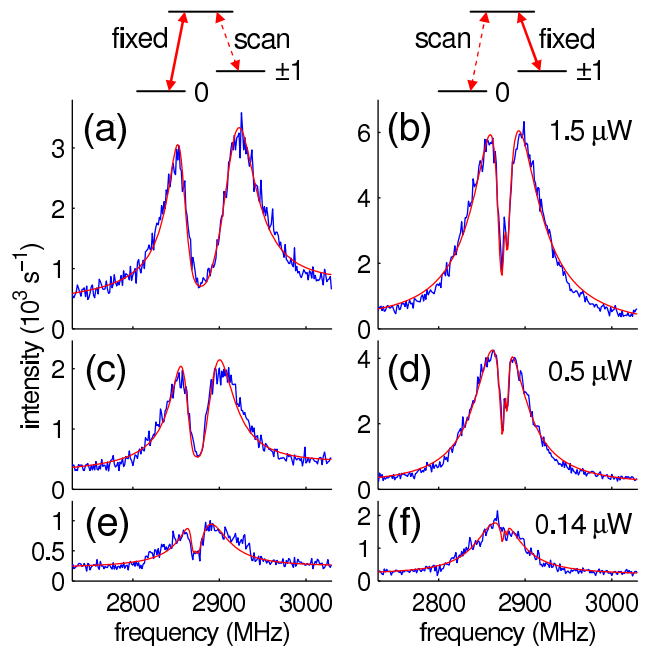}
\caption{\label{fig3} Measured and fitted fluorescence intensity vs. modulation frequency, using the middle excited state of an NV with excited-state level spacings (see Fig. 1) $a=0.31\,\mathrm{GHz}$ and $b=0.94\,\mathrm{GHz}$.  (a,c,e): laser fixed on $m_s=0$ transition; (b,d,f): laser fixed on $m_s=\pm 1$ transition.  Excitation powers: (a,b): $1.5 \, \mu\mathrm{W}$; (c,d): $0.5 \, \mu\mathrm{W}$; (e,f): $0.14 \, \mu\mathrm{W}$.}
\end{figure}
To test agreement between experiment and theory and to obtain information on the individual transition strengths, we performed measurements under varied excitation conditions on another NV center.  For the graphs on the left side of Fig.~3, the laser was on resonance with an $m_{s}=0$ transition while a weak modulation sideband (about 2\% relative power) was scanned across the corresponding $m_{s}=\pm 1$ transitions.  In this case the shape of the fluorescence dip is determined mainly by the strength of the $m_{s}=0$ transition. For the graphs on the right, the roles of the laser and sideband were reversed. Results from several excitation powers are shown along with model fits. To reduce the number of free parameters, we used the constraints $\sum\nolimits_{i}\Gamma_{i} = \Gamma = 2\pi\times 13.4 \,\mathrm{MHz}$, $\Gamma_{3} /\Gamma_{2} = \Omega_{3}^{2} /\Omega_{2}^{2}$, $\gamma_{14} =\gamma_{24} =\gamma_{34} =\Gamma/2+\gamma_{4}$, and $\gamma_{12} =\gamma_{13} =\gamma_{23} =\gamma_{1}$.  For each left/right pair of measurements, the ratios between $\Omega_{i}^{2}$ and the relevant excitation powers were held constant.  The fits in Fig.~3 used $\Gamma_1/\Gamma \approx 0.8$ and $\gamma_4 \approx 2\pi \times 23 \, \mathrm{MHz}$ and also included an adjustable background with linear slope.  Under these conditions we obtain excellent agreement between theory and experiment.  From the fits we can estimate the relative transition strengths for equal excitation power: $\Omega_{2}^{2} /\Omega_{1}^{2} =0.14$ and $\Omega_{3}^{2} /\Omega_{1}^{2} =0.05$.  Thus all ground states are coupled to the excited state. The $m_{s} =\pm 1$ splitting of $5\,\mathrm{MHz}$ visible in the graphs on the right could be due either to strain or to a background magnetic field. We estimate an effective ground-state decoherence rate $\gamma_{1} = 2\pi\times 1.2\,\mathrm{MHz}$, which likely includes optically-induced decoherence mechanisms not described explicitly in the model.  Analysis of the fitted density matrix close to two-photon resonance in Fig.~3a shows a nearly equal statistical mixture of the two dark states. The ground-state coherences $\rho_{12}$ and $\rho_{13} $ are approximately 99\% and 90\%, respectively, of the maximum possible for such a mixture.

To obtain a definite spin superposition, we can apply a weak magnetic field to remove the near-degeneracy of the $m_{s} =\pm 1$ levels. Fig. 4 shows coherent population trapping data on a single NV center for several values of magnetic field. The fluorescence dip appears first near $2.88\,\mathrm{GHz}$ modulation frequency for zero magnetic field and then splits into two dips following the Zeeman splitting of the $m_{s} =\pm 1$ states.  Each dip corresponds to formation of a single dark state. We expect that increasing the magnetic field much further would lead to a decrease in the fluorescence intensity.  Once the $m_{s} =\pm 1$ splitting is larger than the optical transition linewidth, most of the population will collect in the unused ground level.
\begin{figure}[tb]
\includegraphics[]{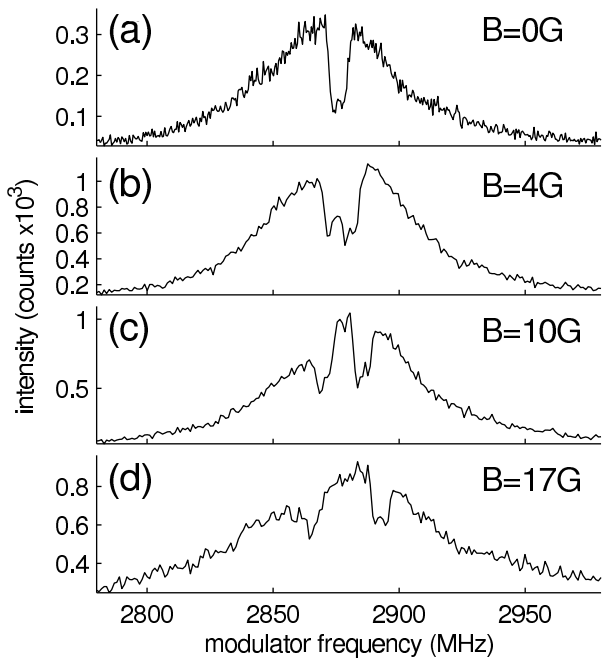}
\caption{\label{fig4} Averaged fluorescence intensity vs. modulation frequency with laser fixed on $m_s= \pm 1$ transitions for various magnetic fields: (a) $0 \, {\rm G}$ (b) $4 \, {\rm G}$ (c) $10 \, {\rm G}$ (d) $17 \, {\rm G}$.}
\end{figure}

These results demonstrate steady-state formation of coherent superpositions of spin states of a single NV center with probability amplitudes directly tunable through the laser and sideband amplitudes.  We expect based on these results that dynamic, optical control of the spin state will also be possible.  Furthermore, this system allows for a tuning of the relative transition strength of the spin-flip versus spin-conserving optical transitions via external parameters such as strain or electric field.  This might prove useful as a way to modulate the photon-spin interaction.  For potential applications, a remaining difficulty is how to reduce the spectral instability and inhomogeneity of the NV centers.  It has recently been demonstrated that individual NV centers can be tuned in frequency through an applied electric field~\cite{tamarat06}, providing a possible solution to this problem.  To fully realize the potential of diamond NV centers for photonic quantum information processing it will be necessary to couple them efficiently to optical cavities and waveguides.  This would enhance the modest oscillator strength of the zero-phonon line and improve the extraction of emitted photons.  Then, schemes for efficient interconversion between photonic and spin qubits could be realized~\cite{fattal06}, and photons could serve as a communications bus between spatially separated spin qubits.  Significant progress has been made toward fabrication of the necessary structures in diamond~\cite{olivero05}.

We thank Nobuhiko Kobayashi for help with XRD measurements.
This work was supported by
DARPA and the Air Force Office of Scientific Research through
AFOSR contract no.\ FA9550-05-C-0017,
DFG (project SFB/TR 21),
EU (Integrated Project Qubit Applications QAP funded by the
IST directorate as Contract Number 015848),
`Landesstiftung B-W' (project `Atomoptik'),
DARPA QuIST under AFOSR contract number C02-00060,
the Australian Research Council,
the Australian Government,
the US National Security Agency (NSA),
Advanced Research and Development Activity (ARDA) and
the Army Research Office (ARO) under contracts
nos.\ W911NF-04-1-0290 and W911NF-05-1-0284,
and the Alexander von Humboldt Foundation.

\end{document}